\documentstyle[epsf,12pt]{article}

\begin{document}

\begin{center}
{\Large {\bf
Gravitationally--Induced Three--Flavor Neutrino Oscillations
as a Possible Solution to the Solar Neutrino Problem}} \\
\vspace{1mm}
J.\ R.\ Mureika\footnote{newt@avatar.uwaterloo.ca}~and~R.\ B.\
Mann\footnote{mann@avatar.uwaterloo.ca}\\
Dept. of Physics, University of Waterloo, Waterloo, ON~N2L 3G1
\end{center}
\begin{abstract}
\small{Neutrinos can undergo flavor--oscillations if they
possess flavor--dependent couplings to the surrounding gravitational
field (the VEP mechanism).  The neutrino fields can be massless,
in accord with the Minimal Standard Model, but at the
expense of the Einstein
Equivalence Principle.  We show that it is possible to explain the
observed Solar Neutrino data from the various experiments using
the VEP solution in a realistic three--generation framework, and
further note how the three--flavor model can offer larger
allowed regions of parameter space over the two--flavor models.}
\end{abstract}

\section{Introduction}
	Various explanations of the Solar Neutrino Problem (SNP) are
based on the assumption that neutrinos possess two non--degenerate eigenbases
in which they can be described.  One of these is the flavor eigenbasis
$|\nu\rangle_W$, relevant to electroweak phenomena, while
the other is diagonal in the quantum mechanical equations of motion.
The first of such models was proposed by Mikheyev, Smirnov, and
Wolfenstein \cite{msw,wolf}, and is therefore dubbed the
MSW mechanism.  In this model, the second eigenstate must be
a massive one, which implies that the neutrinos must have
non--trivial masses.  Such a solution saves the Standard Solar
Model \cite{bah2}, but requires an extension
to the Minimal Standard Model of Particle Physics, {\it i.e.}
massive neutrinos.

	This compromise of the Standard Model can be saved, as
was pointed out in 1988 \cite{gasp}.  Instead of having mass, if
each neutrino couples differently to the (solar) gravitational field
$\phi(r)$, then the same
oscillation mechanism can be obtained (up to the form of the
energy dependence, the primary distinction between the massive
and gravitational oscillation models).  That is, each eigenstate $\nu_i$
has a different coupling $G_i = (1+f_i) G$, with G Newton's
constant, and each $f_i \ll 1$ a dimensionless ``violation
parameter''.  For first generation neutrinos, we take $f_i = 0$,
{\it i.e.} $G_1 = G$.

	The two eigenbases are related by a matrix $V_{N_g} \in SU(N_g)$.
For three flavors, $N_g=3$; discounting CP violations in the
neutrino sector, $V_3$ becomes an orthogonal three--parameter
($\theta_{12},\theta_{13},\theta_{23}$) matrix.  With
$|\nu\rangle_W = V_3 |\nu\rangle_{M,G}$, the neutrino states
evolve according to the equations of motion

\begin{eqnarray}
 & i \frac{d}{dr}\, |\nu\rangle_{M,G} = H_{M,G} |\nu\rangle_{M,G} \nonumber \\
\Longrightarrow & i \frac{d}{dr}\, |\nu\rangle_{W} = H'_{M,G} |\nu\rangle_{W}
\end{eqnarray}
Here, $H_{M,G}$ are the diagonal Hamiltonians for MSW and VEP respectively in
the mass/gravitational eigenbasis,

\begin{eqnarray}
H_M& =& \frac{1}{2E}\; diag\left\{m_1^2,m_2^2,m_3^2\right\} \\
H_G& =& 2E|\phi(r)|\; diag\left\{f_1,f_2,f_3\right\}
\end{eqnarray}
while $H'=V_3^{-1}HV_3+A$ is the
corrected version for the electroweak interactions ($A = diag
\left\{\sqrt{2}G_F N_e,0,0
\right\}$).
It is the off--diagonal nature of $H'$ which induces flavor oscillations,
and the presence of $A$ creates parameter--dependent resonances.
We can subtract a total factor of unity
{}~${\sf{'}}{\hspace{-0.8mm}|\hspace{-0.6mm}|}\hspace{-1.6mm}
\rule[-1mm] {1.5mm}{.1mm}\hspace{-1.1mm}\rule[3.0mm]{.65mm}{.1mm}\:\cdot
H_{11}~$
from $H$, since this yields only an unobservable phase, and hence
deal only with eigenvalue differences $2E|\phi(r)|\Delta f_{21,31}$
($\Delta m^2_{21,31}/2E$ for MSW).

	Re--diagonalization of $H'$ by a matter--enhanced
matrix $V_3^m$ creates a new eigenbasis $|\nu\rangle_{MAT.}$ in which we can
describe the evolution, with $|\nu\rangle_{W} = V_3^m |\nu\rangle_{MAT.}$.
For an electron neutrino $\nu_e$ created in the solar core,
the averaged  probability that it reaches the Earth as a $\nu_e$ is
found to be \cite{bald}

\begin{eqnarray}
\langle P(\nu_e \rightarrow \nu_e) \rangle & =& \sum_{i,j=1}^{3}
|(V_3)_{1i}|^2\; |(P_{LZ})_{ij}|^2\; |(V_3^m)_{1j}|^
2 \nonumber \\
 & = & c_{m12}^2 c_{m13}^2 \left\{ (1-P_1)c_{12}^2
 c_{13}^2 + P_1 s_{12}^2 c_{13}^2 \right\} \nonumber \\
& & + s_{m12}^2 c_{m13}^2 \left\{ P_1 (1-P_2) c_{12}^2 c_{13}^2 +
(1-P_1)(1-P_2)
s_{12}^2 c_{13}^2 + P_2 s_{13}^2 \right\} \nonumber \\
& & + s_{m13}^2 \left\{ P_1 P_2 c_{12}^2 c_{13}^2 + P_2 (1-P_1) s_{12}^2
s_{13}^2 +(1-P_2) s_{13}^2 \right\}~. \nonumber \\
{  }
\end{eqnarray}
Here, $s_{ij},c_{ij} \equiv \sin{\theta_{ij}},\cos{\theta_{ij}}$ and
$s_{mij},c_{mij} \equiv \sin{\theta_{ij}^m},\cos{\theta_{ij}^m}$ are
the parameters of $V_3$ and $V_3^m$, respectively.  The matrix
$P_{LZ}$ has elements $P_{1,2}$ which describe the probability of
non--adiabatic level crossing between $\nu_e \rightarrow \nu_{\mu}~,
{}~\nu_e \rightarrow \nu_{\tau}$ (hereafter 12-- and 13--transitions).

\section{Behavior of $\langle P(\nu_e \rightarrow \nu_e) \rangle$ for a Double
Resonance}

If the $\nu_e$s are created at electron densities higher than the
corresponding resonance density for either 12-- or 13--transitions,
then the matter--enhanced mixing angles approach the value
$\theta_{12}^m,\theta_{13}^m \rightarrow \frac{\pi}{2}$.  Hence,
the survival probability reduces to the simpler form \cite{jrm1}

\begin{equation}
\langle P(\nu_e \rightarrow \nu_e) \rangle = P_1 P_2 c_{12}^2 c_{13}^2 + P_2
(1-P_1)
s_{12}^2 s_{13}^2 +(1-P_2) s_{13}^2~.
\end{equation}
To see how this is affected by the addition of the third
flavor $\nu_{\tau}$, we can examine its limiting form
for small and large $\theta_{13}$.  In the former case,
we have

\begin{equation}
\langle P(\nu_e \rightarrow \nu_e) \rangle = c_{12}^2 P_1 P_2~,
\end{equation}
which shows energy dependence through both 12-- and 13--transitions
in the $P_{1,2}$ terms.  Solutions to the solar neutrino problem
in this case are similar to the small--angle solution in the
two--flavor limit, and are of questionable statistical validity
\cite{jrm1}.  However, we note that for large $\theta_{13}$, the
term $P_2 \rightarrow 0$ (adiabatic approximation for 13--transition),
and so the above expression further reduces to

\begin{equation}
\langle P(\nu_e \rightarrow \nu_e) \rangle = s_{13}^2~.
\end{equation}

\begin{figure}
\begin{center}
\leavevmode
\epsfysize=200pt
\epsfbox[-80 80 740 710] {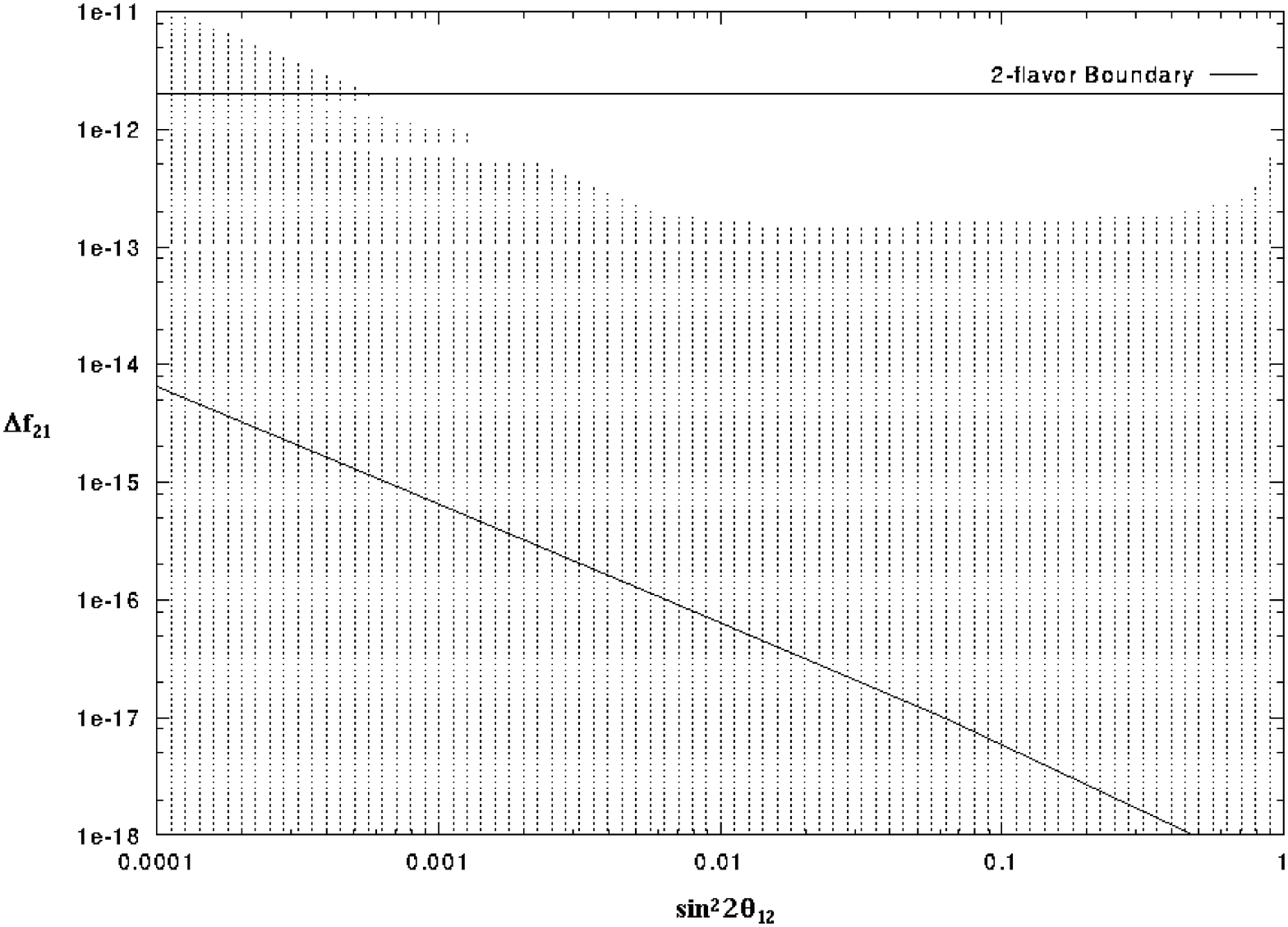}
\end{center}
\caption{$3\sigma$ overlap for data from [3], with $\Delta f_{31} =
10^{-13}~,~s_{13}^2 = 0.4$}
\label{snu1}
\end{figure}

	This limiting form has interesting implications, as it
suggests that the $\nu_e$ suppression is not only
energy--independent (as is usually the case with large--angle
oscillation solutions), but that it is also independent of
the 12--transition.  Figure~\ref{snu1} shows the allowed
parameter--space overlap for the most recent solar neutrino
experiment data in the large $\theta_{13}$ case.  Clearly,
the addition of a third flavor greatly broadens the regions
from the much smaller two--flavor results (see \cite{bah1}
for these).

\section{Comparison of ${^8}\!B$ Neutrino Fluxes}

\begin{figure}
\begin{center}
\leavevmode
\epsfysize=200pt
\epsfbox[-135 90 730 695] {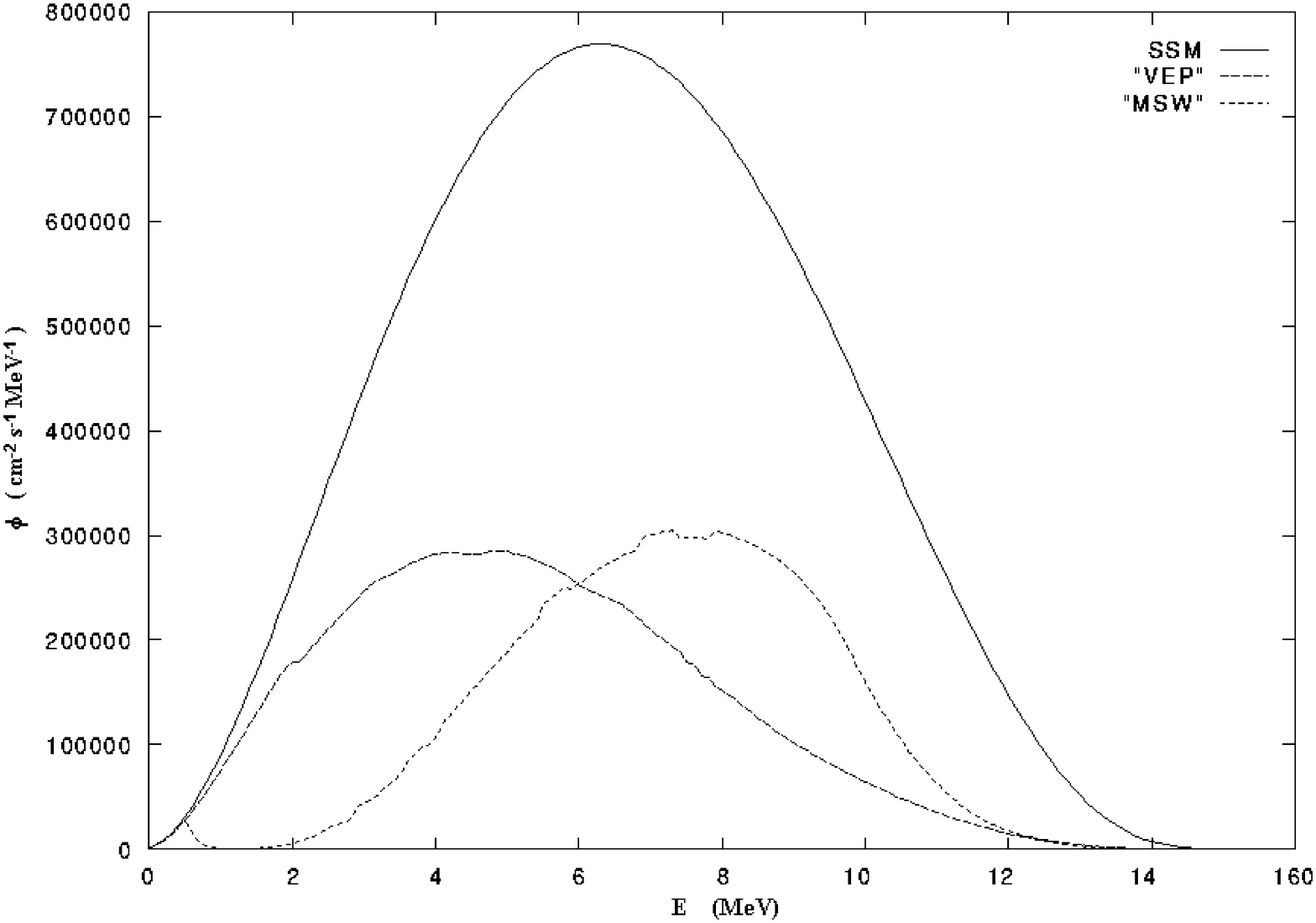}
\end{center}
\caption{MSW and VEP reduced fluxes yielding counting rate $R = 2.00~$SNU,
with both 12-- and 13--resonances allowed.}
\label{flux1}
\end{figure}

	As previously mentioned, the difference between the two
oscillation mechanisms resides in their energy dependence.  A study
of the spectrum of ${^8}\!B$ neutrinos incident on terrestrial
detectors can help shed light
on which suppression mechanism, if any, is at work.  In the
previous section, the large $\theta_{13}$ form was shown
to be energy independent, while the small $\theta_{13}$ case
is a function of both 12-- and 13--transitions.  So, we should
expect to see some type of spectral distortion in the
${^8}\!B$ neutrino flux in this small angle case \cite{jrm1,jrm2}.
Figure~\ref{flux1} shows that the attenuation of $\nu_e$s
is indeed affected
quite differently by each model: MSW suppresses low--energy
neutrinos, while VEP suppresses higher--energy ones.  Since
detectors such as Kamiokande II, or SNO
(when it comes online) can detect subtle variations in the
high energy portion of the flux spectrum, we would expect
such behavior as that in figure~\ref{flux1} to be a major
clue as to the solution of the SNP.
\vskip .25 cm

\noindent
{\bf Acknowledgements}

This work was supported in part by the Natural Sciences and Engineering
Research Council of Canada.

\pagebreak

\end{document}